\documentclass{article}

\usepackage{epsf}
\usepackage{graphicx}
\usepackage{amssymb,ComplexSystems}
\usepackage{amsmath}
\usepackage{amssymb}

\begin{document}

\title{Definition and Identification of Information Storage and Processing Capabilities as Possible Markers for Turing-universality in Cellular Automata%
%
}

\author{\authname{Yanbo Zhang}\\[2pt]
\authadd{Physical Department, University of Science and Technology of China}\\
\authadd{Hefei, Anhui, P.\ R.~China}\\
\and
}

%
%
\markboth{Complex Systems}
{}

\maketitle

\begin{abstract}
To identify potential universal cellular automata, a method is developed to
measure information processing capacity of elementary cellular automata.
We consider two features of cellular automata: Ability to store information,
and ability to process information. We define local collections of
cells as particles of cellular automata and consider information
contained by particles. By using this method, information channels
and channels' intersections can be shown. By observing these two features,
potential universal cellular automata are classified into
a certain class, and all elementary cellular automata can be classified
into four groups, which correspond to S. Wolfram's four classes:
1) Homogeneous; 2) Regular; 3) Chaotic and 4) Complex. This result
shows that using abilities of store and processing information to
characterize complex systems is effective and succinct. And it is
found that these abilities are capable of quantifying the complexity
of systems.
\end{abstract}


\section{Introduction}
\label{intro}
A universal system is a system that can execute any computer program.
In other words, it is feasible for it to execute any algorithm~\cite{banks_universality_1970}.
It is found that some systems with simple rules can be a universal
system, such as rule~110 in elementary cellular automata~\cite{cook_universality_2004,banks_universality_1970, wolfram_new_2002}.
Some tag systems and cyclic tag systems are also proved to be universal,
which are also systems with simple rules~\cite{cook_universality_2004,wolfram_new_2002,cocke_universality_1964}.
Glider system, which is an idealized system to simulate particle process
of real physics system, was also proved to be a universal system~\cite{cook_universality_2004}.
And particle machines in periodic backgrounds was proved to
be universal~\cite{jakubowski_when_1996}.

The widespread existence of universal systems implies that some process with simple rules in the real world may be able to execute some algorithms or any algorithm. Because of the significant amount of algorithms, these systems' behaviors can be changeful and complex, which was considered as a potential origin of complexity in~\cite{wolfram_new_2002, cross_boundary_Jurgen_Zenil_2015}.

For cellular automata can show the wide variety of complex phenomena 
in the real world, and cellular automata are also sufficient
generality for a wide variety of physical, chemical, biological,
and other systems~\cite{wolfram_statistical_1983}. Identifying
universal cellular automata will help people understand origins of
cellular automata's behaviors and find key dynamics of computation.

In this study, a method is developed to identify potential universal elementary
cellular automata. Two abilities of a system are considered:
1) Ability to store information and 2) Ability to process information.
We found these two features can identify potential universal cellular
automata and quantify the complexity of systems.

\subsection{Elementary Cellular Automata}

Cellular Automata (CA for singular, CAs for plural) are ideal models
for physical systems in which space and time are discrete. And elementary
cellular automata (ECA for singular, ECAs for plural) is one of the
simplest kind of CAs.

ECAs are dynamic systems defined by deterministic rules, working on
a 1-dimension list $\{c_{n}\}$ with $n$ cells. Rules can be expressed
by function $F$:
\begin{equation}
c_{n}(t+1)=F[c_{n-1}(t),\, c_{n}(t),\, c_{n+1}(t)],
\end{equation}
where $n\in\mathbb{Z}$.

Therefore, $c_{n}(t+1)$ is the function of itself $c_{n}(t)$ and
its two immediate neighbors: $c_{n-1}(t)$ and $c_{n+1}(t)$. Each
$c_{n}\left(t\right)$ has two possible states,~$0$ or~$1$. So
there should be a~$2^{3}=8$ length list $R$ to define a rule, and
there will be~$2^{8}=256$ different rules. When $R$ is equal to
$\{0,0,0,1,1,1,1,0\}$, by considering it as a binary code, it will
equal to~$30$ in decimal base,
which is the ECA rule~30.

With a given initial list $L_{0}$, an ECA will apply the function
$F$ to all cells parallelly to update $L_{t}$ to $L_{t+1}$. i.e.,
\begin{equation}
L_{t}\xrightarrow[\,]{F}L_{t+1}.
\end{equation}

\begin{figure}
\begin{centering}
\includegraphics[width=8.5cm]{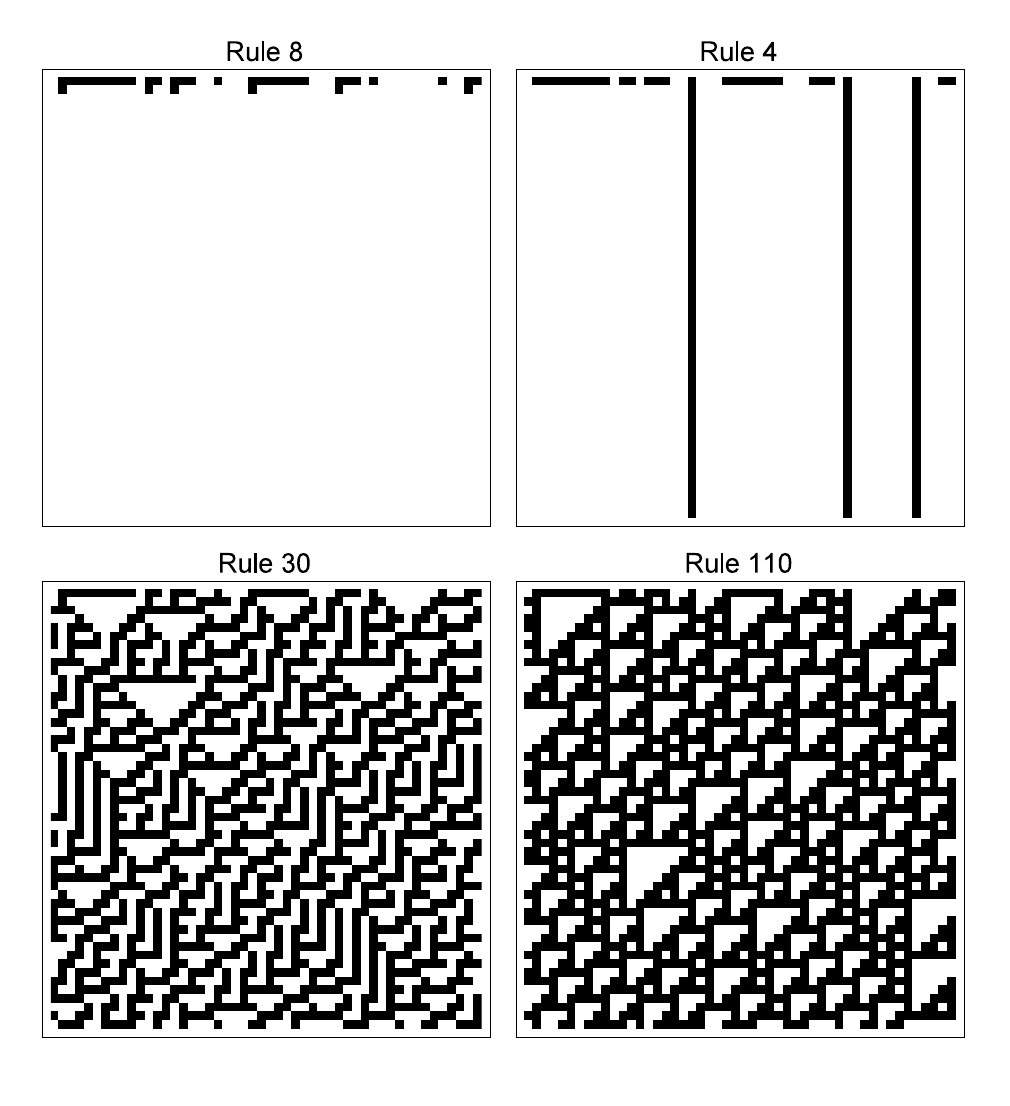}
\par\end{centering}

\protect\caption{\label{fig:Four-classes}Evolution of~4 typical rules from class~1$\sim$4.
Rule~8 is in class~1, rule~4 is in class~2, rule~30 is in class~3,
and rule~110 is in class~4.}
\end{figure}

By doing this process repeatedly, a matrix $M^{(\text{rule})}=\left(L_{0},L_{1},\ldots,L_{t}\right)$
will be generated, which is the ``space--time evolution''. Figure~\ref{fig:Four-classes} shows two space--time evolutions
generated by ECA rule~30 and ECA rule~110, started with the same $L_{0}$.


256 different ECAs can be classified. In this paper, we compare our work with Wolfram's classification, which are class~1$\sim$4  in~\cite{wolfram_new_2002,wolfram_statistical_1983}.
 The classes are: 1) Homogeneous; 2) Regular; 3) Chaotic; 4) Complex.
Some typical space--time evolutions are shown in Figure~\ref{fig:Four-classes}. There are also some other classifications, see~\cite{note_CA_Class_2013, asymptotic_in_CA_2013, compression_class_CA}.

\section{Methodology}

We consider two abilities of ECA rules: Ability to store information,
and ability to process information. The ability to store information
will make the system stable enough and do not have too much noise.
Only when information can be stored, information can move stably in
a system, so that the whole system can be related. The ability to
process information means interactions between information should
be found in a system.


We define a system can store information when its current local states 
can be used to infer previous states at some location. It's true that some reversible 
systems can store all information at the whole system, but this information can 
hardly be used to infer the previous states because many of them are computational irreducible. 
Thus, the particle systems can cover the definition.

We identify potential universal ECAs based on a theorem proposed in
\cite{jakubowski_when_1996}, which considers particle-like structures
and their behavior in systems to identify Turing machines and UTM.

A method was developed to extract particle patterns from ECAs to build
``particle machines'', and to measure their computation ability
by taking into account their features. First, it is necessary to introduce
particles machines and define particles in ECAs.

\begin{figure}
\begin{centering}
\includegraphics[width=8.5cm]{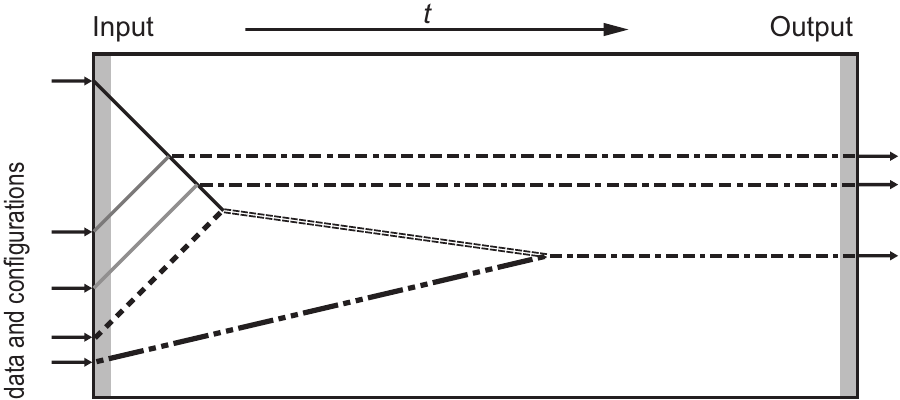}
\par\end{centering}

\protect\caption{\label{fig:Data-and-configurations}A typical particle machine.}
\end{figure}

\subsection{Particle machines}

A particles machine~(PM), is a system in which particles
can move, collide, annihilate and generate in a homogeneous medium.
Figure~\ref{fig:Data-and-configurations} shows a typical PM. Data
and configurations are injected from left in the form of particles,
and by executing this system, particles will have interactions. Lines
and dotted lines in this figure represent the paths of particles.
After time $t$, the system will generate an output. The identity
of a particle includes position, phase, and velocity. During collisions,
particles can alter their identities, or be generated or annihilated.
These changes of particles can be considered as a function of particles
that participate in the collision, which is the collision function.
Some particles machines are proved to be Turing machines or universal
Turing machine (UTM) in~\cite{jakubowski_when_1996}. A PM is at
least a Turing machine when: 1) Identity of particles can change
during collisions; 2) Collision function is depending on identities
of particles. For the first requirement, the identity of particles can
change during collisions, also means new particles can be generated
 during collisions. And the second requirement means the result of
a collision should depend on types of particles that participate in
the collision. If no particles can be generated or annihilated in
collisions in a PM, then the PM is not a UTM.

\subsection{Particles in ECAs}

We define a local grid of cells in $M^{(\text{rule})}$ as a particle in ECAs. Here
we consider one kind of particles: Their sequence may change periodically
or not change through time. We call them ``elementary particles''.
It will be practical if we start with these simple kind of particles.

Particles contain information, so that information can move in
space, and have interaction with other information, which is a kind
of computation~\cite{collision_based_computing_Adamatizky_2012}. All identities of particles: Location, velocity, and
sequence, can be computed by collisions. And all of these identities
can be preserved if there are no collisions.

\begin{figure}
\begin{centering}
\includegraphics[width=8.5cm]{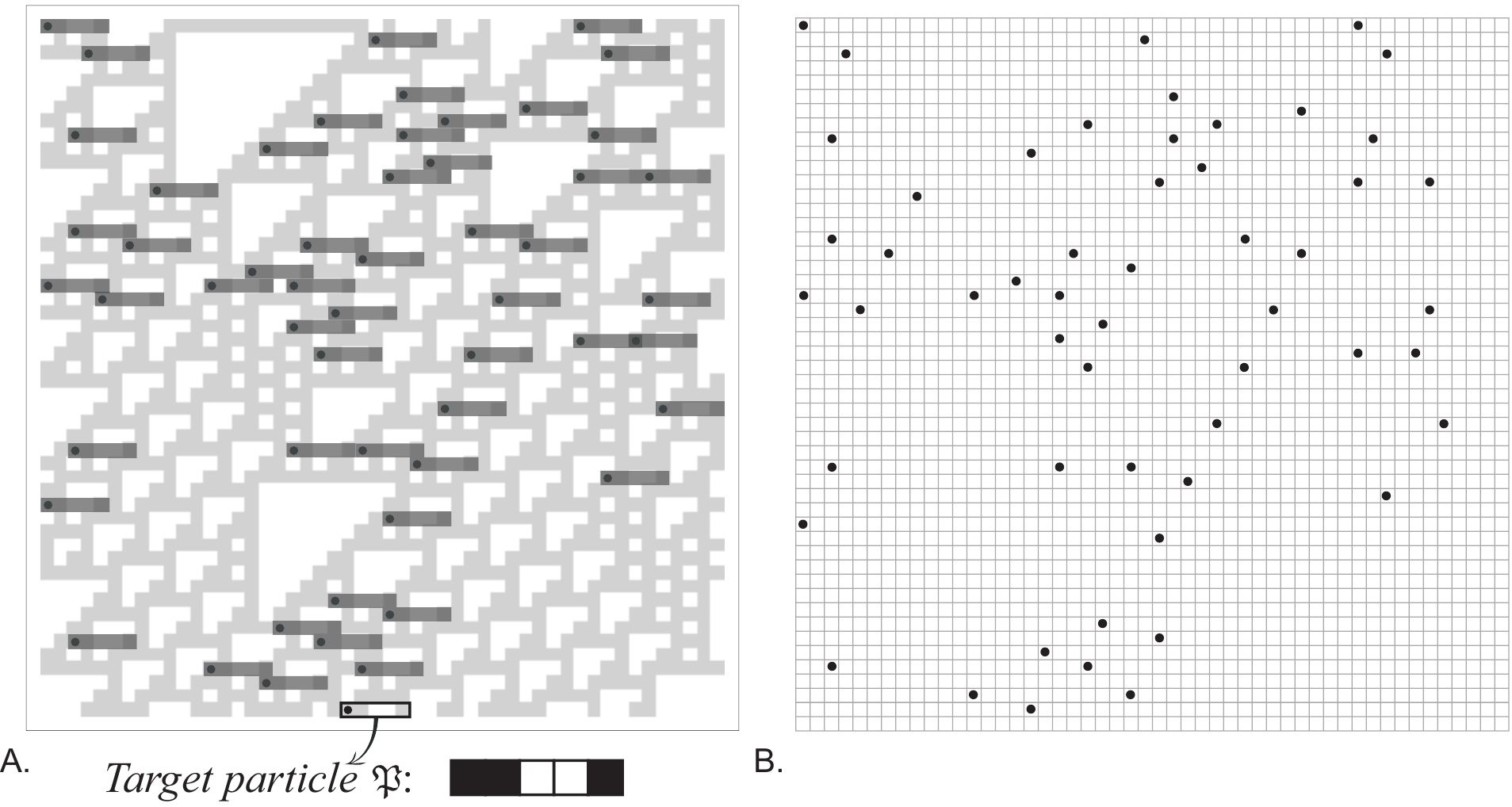}
\par\end{centering}

\protect\caption{\label{fig:This-figure-shows}\textbf{A)}. An illustration of how
it takes a ``target particle'' $\mathfrak{P}$ from a matrix generated
by ECA rule~110. The rectangle with black frame is the target particle
$\mathfrak{P}$, and gray rectangles mean there has a similar sequence
as~$\mathfrak{P}$, which are linear particles $\mathfrak{L}$s. In
this figure, the $\mathfrak{P}$ and $\mathfrak{L}$s are \protect\includegraphics[height=0.2cm]{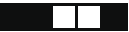},
and found~$54$ same sequences. \textbf{B)}. A figure of matrix $\mathfrak{M}^{(\text{110})}$. $\mathfrak{M}^{(\text{110})}_{t,x}=p_0$ when there is a $\mathfrak{L}$ at $\{t,x\}$, or $\mathfrak{L}_{t,x}=0$. Dots at $\{t,x\}$ means $\mathfrak{M}^{(\text{110})}_{t,x}=p_0$.}
\end{figure}

To extract particles' identities from ECAs' space--time evolutions, a certain
sequence should be chosen for the research. We need to choose a sequence
as a particle to study, which is the \textquotedblleft target particle\textquotedblright .
As Figure~\ref{fig:This-figure-shows}.A shows, we choose target
particle $\mathfrak{P}$ at the center-bottom of a space--time evolution 
and mark the same sequences as ``linear particles'' $\mathfrak{L}$s, which are 
the dots in~Figure~\ref{fig:This-figure-shows}.B.

$\mathfrak{P}$ can be explained by the equation $\mathfrak{P}=L_{t_{\text{max}}}(p_{L},p_{R})$, where 
$L_t$ is the $t$--th row of the space--time evolution. $p_{L}$ and $p_{R}$ is the start-index
and the end-index for $\mathfrak{P}$.

A particle $P$ at $\left(t;\overrightarrow{x}\right)$, its location
may be $\left(t';\overrightarrow{x}'\right)$ at time $t'$~($t'<t$).
We call the particle at $\left(t';\overrightarrow{x}'\right)$ as
$P$'s father-particle $P_{f}$. If let $P$ be $\mathfrak{P}$, the
$P_{f}$ will be one of the $\mathfrak{L}$s.

All $\mathfrak{L}$s in $P$'s light cone are possible to be the father-particle of $\mathfrak{P}$
(i.e. $\mathfrak{P}_{f}$), we assume that there is one and only one
$\mathfrak{L}$ is the $\mathfrak{P}_{f}$, and each $\mathfrak{L}$
has probability $p$ to be the $\mathfrak{P}_{f}$. So when there
are $n$~$\mathfrak{L}$s, the probability~(i.e.\ $p_{0}$) for a~$\mathfrak{L}_{i}$ to be a father-particle is:

\begin{equation}
p_{0}(p,n)=p\left(1-p\right)^{n-1}\,.
\end{equation}

All the $\mathfrak{L}_{i}$ are drawn on a matrix $\mathfrak{M}^{(\text{rule})}$,
such as Figure~\ref{fig:This-figure-shows}.B. $\mathfrak{M}^{(\text{rule})}_{t,x}=p_0$ when there is a $\mathfrak{L}$ at $\{t,x\}$, or $\mathfrak{L}_{t,x}=0$.
We call $\mathfrak{M}^{(\text{rule})}$ ``probability matrix''. The positions with black points will add
a number $p_{0}$. Each black point means there is a linear particle
$\mathfrak{L}$ of $\mathfrak{P}$ at~$\left(t,x\right)$,~$(t,x)$
is the location of the black point.~$\mathfrak{M}_{t,x}$ equals to~$p_{0}\left(p,n\right)$.

\begin{figure}
\begin{centering}
\includegraphics[width=10cm]{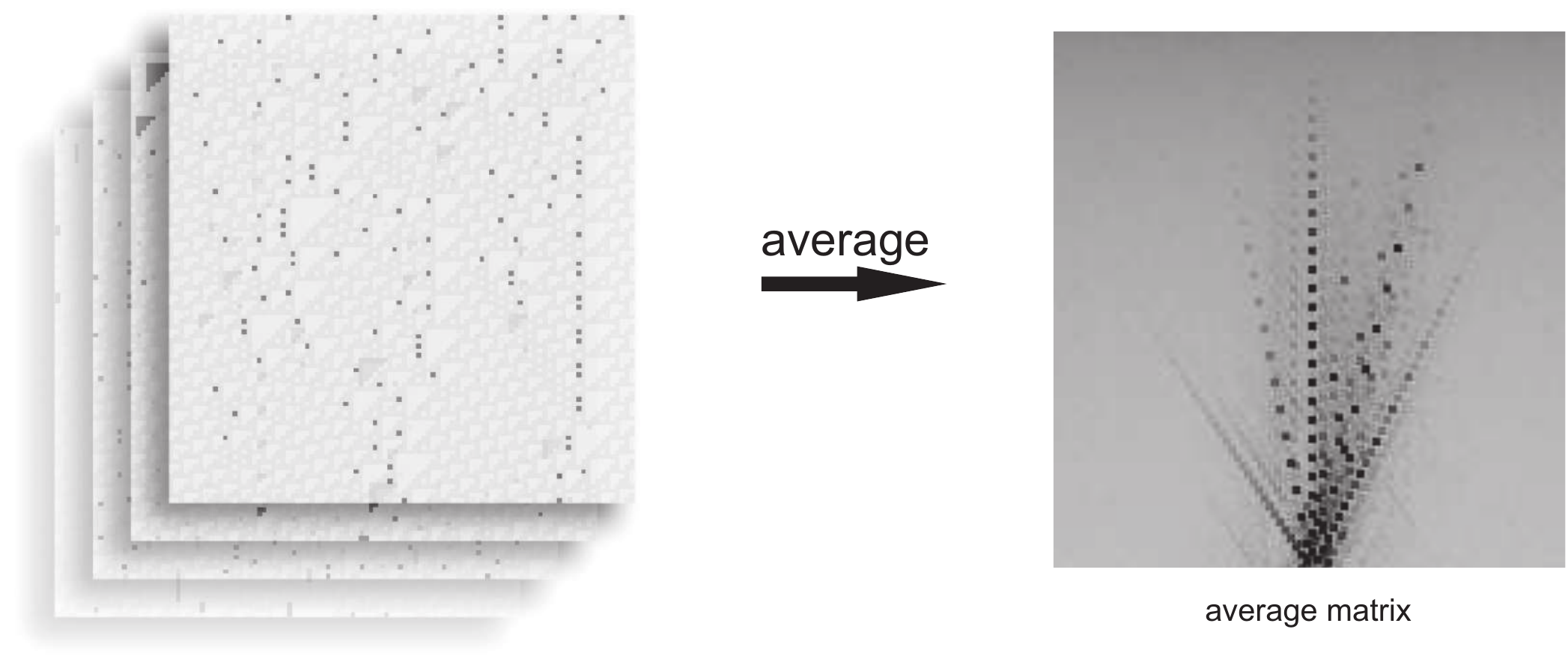}
\par\end{centering}

\protect\caption{\label{fig:averageM}The average matrix $\overline{\mathfrak{M}}^{(\text{110})}$,
generated with~$10^{6}$ probability matrices, with $p$ equal to~$0.01$ for Equation~(1). }
\end{figure}

The average $\mathfrak{M}^{(\text{rule})}$ that generated with random initial lists:

\begin{equation}
\overline{\mathfrak{M}}^{(\textrm{rule})}=\frac{1}{N}\sum_{i=1}^{N}\mathfrak{M}_{\textrm{random}}^{(\textrm{rule})}\,,
\end{equation}

\noindent will show some patterns that represent particles and particles'
behavior. We call~$\overline{\mathfrak{M}}^{(\textrm{rule})}$ ``average
matrix''. Figure~\ref{fig:averageM} shows how an average matrix
was generated.

The meaning of an average matrix is, if a particle is found at the
center-bottom of a space--time evolution, it may come from position~$(t,x)$
with probability~$\overline{\mathfrak{M}}_{t,x}/\sum_{t,x}\overline{\mathfrak{M}}_{t.x}$.
So the pattern in an average matrix represents traces of particles.
We calculate the average matrix with~$N=10^{6}$ for each rule.

\subsection{Extracting particle's identity from an average matrix}

By observing patterns of average matrices, the identity of particles can
be extracted. A typical average matrix is shown in Figure~\ref{fig:averageM}.
If particles can emerge, there will be some lines in the average matrix.
Each line represents at least one particle, and their variations show
interactions between particles.

The change of a line's intensity with time represents interactions between
particles. Because if a particle is moving straight without any interactions,
the lines' intensity will not change through time. But if the particle
can be generated by other particles, it will not be found before it
was created, so that the intensity will change through time, mostly,
the intensity will get higher when~$t$ is getting higher.

\section{Result}

\begin{figure}
\begin{centering}
\includegraphics[width=10cm]{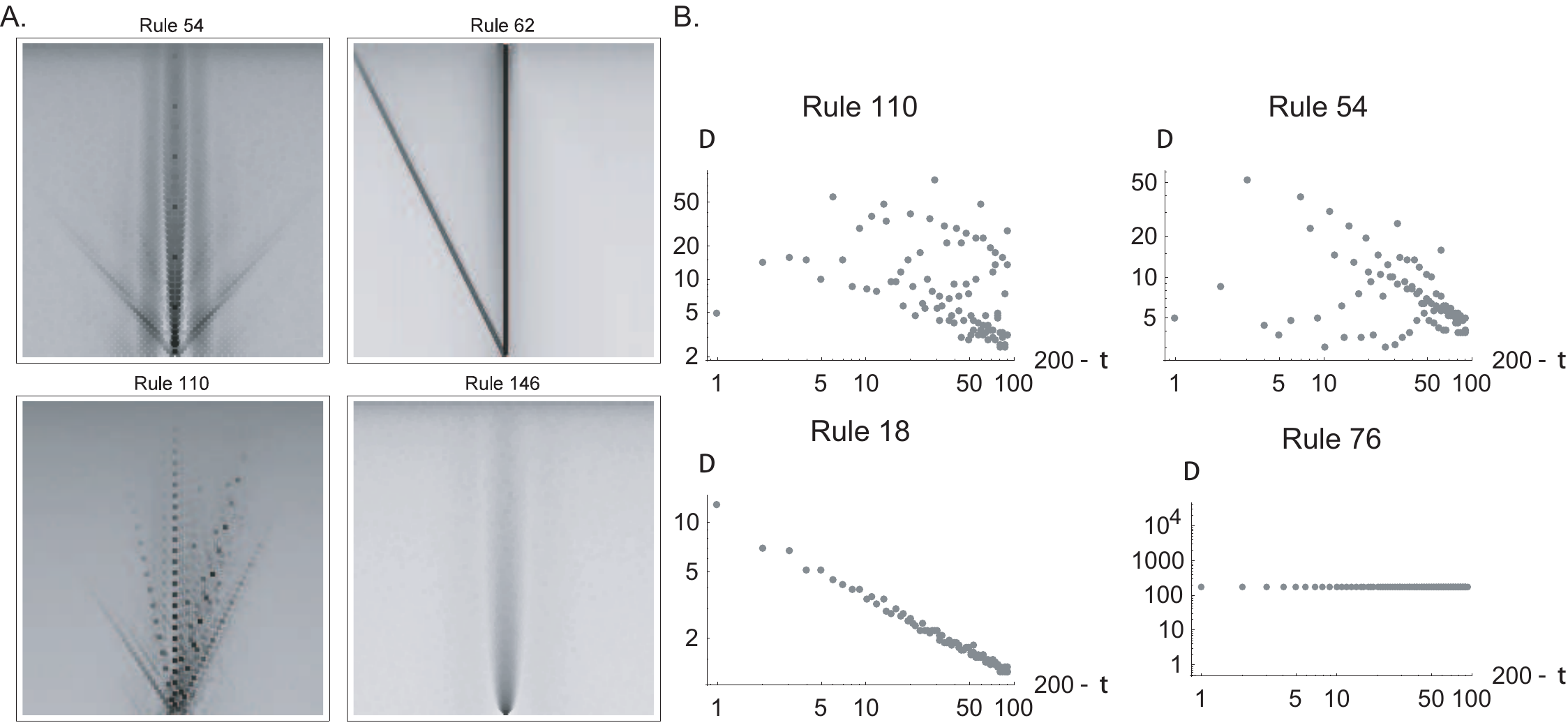}
\par\end{centering}

\protect\caption{\label{fig:typicalRules}\textbf{A)}. Four typical $\overline{\mathfrak{M}}$.
\textbf{B)}. The intensity change with time for four typical rules.
The $t$-axis is time, and $D$-axis is the intensity of particle
traces. It can be seen that the intensity $D(t)$ may change with time
for some rules.}
\end{figure}

We get $\overline{\mathfrak{M}}$ for all rules,
some typical $\overline{\mathfrak{M}}$ shown in Figure~\ref{fig:typicalRules}.

We get numbers of particles' traces for each ECA rules, which correspond
to the number of particles. All traces are straight lines with various
angles. For rules shown in Figure~\ref{fig:typicalRules}, rule~54
has~3 traces, rule~62 has~2 traces, rule~110 has more than~6
traces, and rule~18 has a smooth trace. We use $T(\text{rule})$
to represent the count of traces, such as $T(54)=3$, which can be
used as a parameter to classify ECAs.

The intensity of traces may change through time.
The result shows that they have two kind behaviors: 1) Constant,
2) Variational (mostly, the intensity getting higher when $t$ is getting
higher). We use $C(\text{rule})$ to represent the existence of variation,
such as $C(54)=1$ ($1$ is variational,~$0$ is constant ). These two behaviors
can be used as a parameter to classify ECAs. In Figure~\ref{fig:typicalRules},
traces in rule~54,~62 and~110 are getting more obvious when time
$t$ gets higher. Figure~\ref{fig:typicalRules} shows how the intensity
of particle traces variation with time, where $D(t)=\max(L_{t})$.

Power law show in some rules, where $D(t)\sim(t_{\textrm{max}}-t)^{-\alpha}$,
such as Rule~146 and Rule~18, such power law also found by~\cite{letourneau_particle_2010}~(see Figure~\ref{fig:Particles-146}).

\subsection{Identifying Turing Machines and Potential UTM}

To identify Turing machines and potential UTM, the two parameters we mentioned
above will be used to classify ECA rules into four classes. According
to the theorem of particle machines~\cite{jakubowski_when_1996},
when $T(\textrm{rule})\geq2$ and $C(\textrm{rule})=1$, then this
ECA rule behave as a Turing machine and potentially be a UTM. A particle
machine that is a Turing machine should have at least~$2$ particle
traces so that it is possible to have interactions between particles.
And traces' intensity should change, which represents that new particles
can be generated during collisions. So all rules can be classified into
four classes: A). $T\geq2$ and $C=1$; B).~$T<2$ and $C=1$; C).
$T<2$ and $C=0$; D). $T\geq2$ and $C=0$.

\begin{figure}
\begin{centering}
\includegraphics[width=10cm]{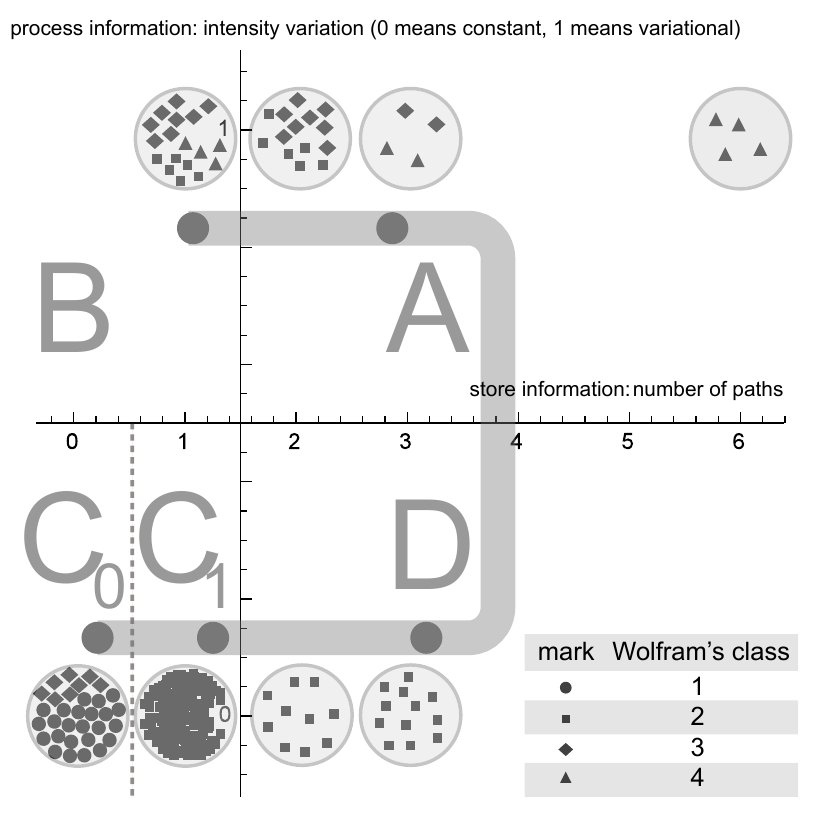}
\par\end{centering}

\protect\caption{\label{fig:class}The classification of ECAs, divided by \emph{the number
of paths} and \emph{intensity variation}. \emph{The number of paths} associated with the ability to store information, and 
\emph{intensity of variation} associated with the ability to processing information. 
In each phase, rules will have similar behaviors. In the phase-A, all rules have both a high
number of traces ($T\geq2$) and interactions that can generate particles,
so that it is possible for these rules to have complex behaviors.
The shape of a point represents its class in Wolfram's classification.
Each point in this figure represents a rule, and their positions were
moved randomly~($\sim 0.3$) so that they can be seen clearly without too many overlaps.}
\end{figure}

Figure~\ref{fig:class} shows the final classification for all rules
of elementary cellular automata. Each point represents a rule for
an elementary cellular automaton. The $x$-axis is ``number of traces'',
and $y$-axis represent the existence of information traces' changes,
where~$0$ means constant,~$1$ means variational. The shape of
a point represents its class in Wolfram's classification.

In class~A, rules have complex behaviors, and many particles with
plentiful interactions can be found. The information here will be stored
and processed. Then they can be considered as a Turing machine with enough complexity and computation ability,
 which was considered to have connections with Turing universality~\cite{cross_boundary_Jurgen_Zenil_2015, Asymptotic_Intrinsic_2016}; 
In class~B, rules will generate some random patterns,
particles have too many interactions with the background, so that information
traces are dissipated. The information here cannot be stored; In class~C,
rules will generate continuous or random structures without any complex
behavior. Rules in this class do not have particles or have particles
but no interactions. In class~D, rules will generate some structures
that do not have enough interactions, which will not have any complex
behavior either. New particles cannot be generated during collisions.

Class~C can be divided into two subclasses, as shown in Figure~\ref{fig:class},
separated by a dotted line. We use ``$\text{Rule}_{x}$'' to express
the subclasses. $\text{C}_{0}$ means the subclass of class~C with
$T$ equal to~$0$. $\text{C}_{1}$ means a subclass of class~C
with $T$ equal to~$1$. In $C_{0}$, rules do not have any particles,
the information here cannot be stored or processed. In $C_{1}$, rules
have particles but do not have interactions between particles. The information
here can only be stored but cannot be processed.

When going through the dark curve in Figure~\ref{fig:class} (anticlockwise),
the frequency of finding interactions is continually growing. And when
the frequency is higher than it in class~A, it will generate too
much noise, so particles and information will be scattered. When it
is lower than the frequency in class A, the number of interactions
is not enough to do computation or universal computation, so the behavior
is too simple to get complex behaviors.

\begin{figure}
\begin{centering}
\includegraphics[width=11cm]{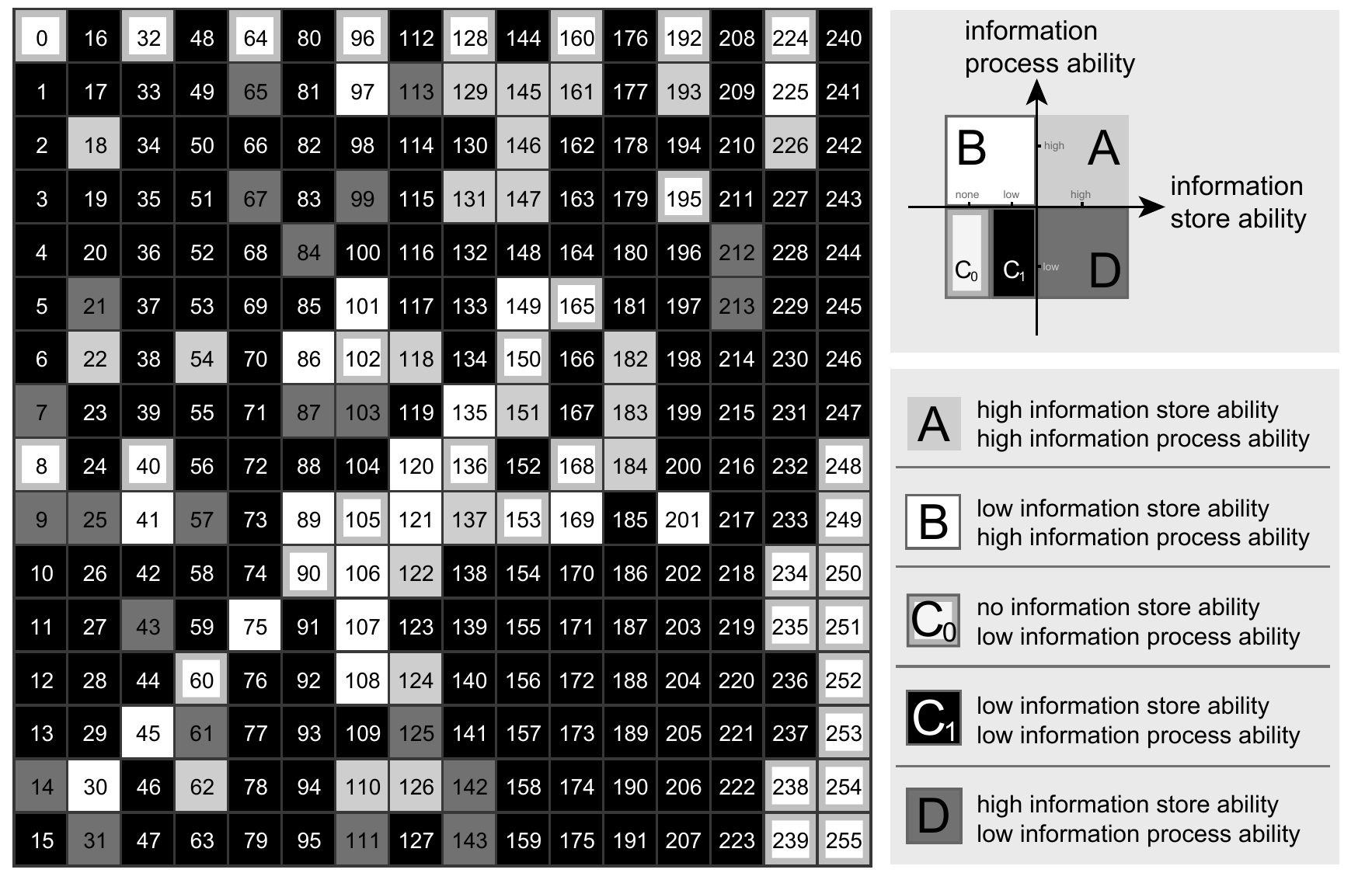}
\par\end{centering}

\protect\caption{\label{fig:classMat}This is the final classification of 
all ECA rules with the method introduced in this paper. 
In this matrix, each kind of texture or color 
represents a class defined by this paper (see the column at right side), and the numbers over each square are the rule indexes. 
Each texture is associated with the ability of processing and storing information, which corresponding to the computation ability. 
The class~A, which have both high information store and process ability, is considered having high computation ability. Rule~110, which is a UTM, is classified into this class.
}
\end{figure}

Some typical rules in these 4 classes show in Table \ref{tab:Typical-Rules-for}. All rules' classification are shown in Figure~\ref{fig:classMat}.

\begin{figure}
\begin{centering}
\includegraphics[width=10cm]{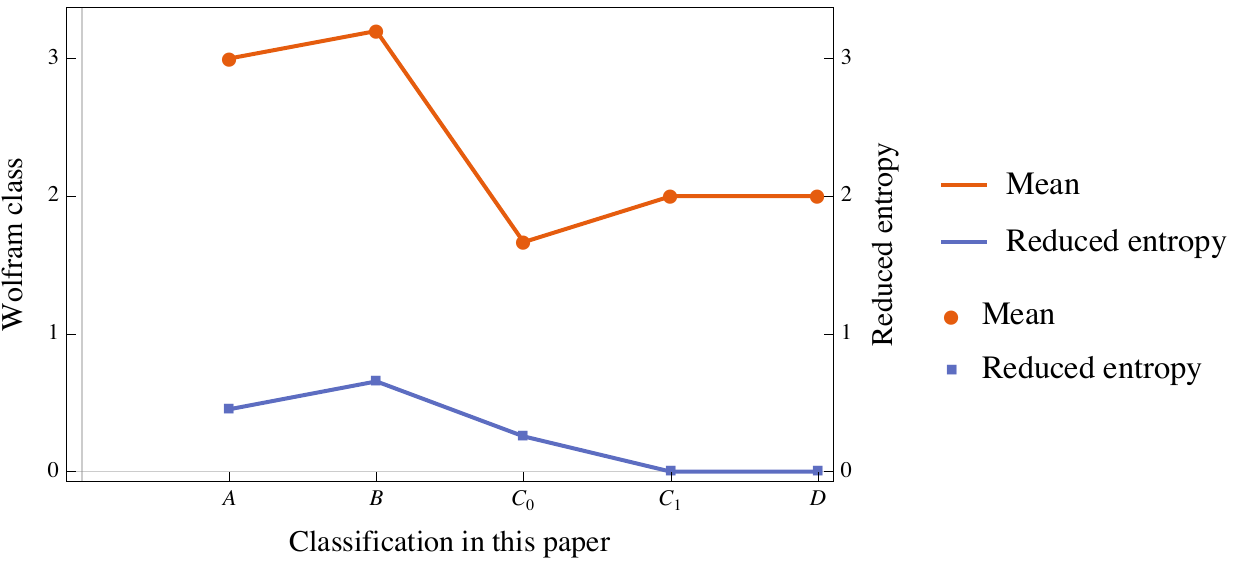}
\par\end{centering}

\protect\caption{\label{fig:img_Compare_PC_WC}
This figure shows the relation between Wolfram's classification 
and the classification in this paper. The orange line with dot markers is the average of $W_i$, 
which is Wolfram's classes of the rules in class $i$ of this paper. The average numbers only 
make sense when all rule in a certain class (of this paper) have a same Wolfram class 
because the Wolfram class do not have an order. The blue line with square markers is 
the reduced entropy of Wolfram classes of rules in certain class in this paper, which can 
measure the correlation between these two kinds of classification. The reduced entropy is 
defined as $h = H/H_{\text {max}} = H / \log n$, where $H$ is the entropy of $W_i$, and $n$ 
is the length of $W_i$.
}
\end{figure}

The relation between this classification and Wolfram's classification 
was also studied. According to Figure~\ref{fig:img_Compare_PC_WC}, 
Class $C_1$ and $D$ have a strong correlation to a certain Wolfram class, which is Class 2. While class 
$A$, $B$, and $C_0$ contain some different Wolfram classes. Here the reduced 
entropy is used to measure the relation between the two classifications because 
the Wolfram classification dose not have an order. The reduced entropy is defined as 
$h = H/H_{\text {max}} = H / \log n$ where $H_{\text {max}}$ is the maximum that entropy $H$ could be.

\section{Discussions}

In this study, we consider two abilities as key dynamics for computation:
\begin{enumerate}
  \item Ability to store information;
  \item Ability to process information.
\end{enumerate}

The ability to store information means there should be particles emerge
in a system so that information can move in the system. And in this
way, the whole system can be connected and linked to be an entirety,
which was considered as a common feature of complex systems. 
Ability to process information means the system can compute information and
execute algorithms. 

By using the coarse-grained method, robust patterns
can be found, rules with different computation abilities are classified into a particular
class~(class~A, shown in Figure~\ref{fig:class}).

All ECA rules can be classified into four classes, which correspond to Wolfram's
classification. All rules in class~1 and most rules in class~2
(Wolfram's classification), was found do not have interactions that
can generate new particles. Most rules in class~3 are found do
not have enough particles to perform the universal computation. All rules in class~4
are found classified into class~A in this study. For rule~146, 183,
18 and 22, which are classified into class~3 (chaotic) by Wolfram,
are classified into class~A in this study, which means these
rules are capable of doing complex computations. This result corresponds to the research~\cite{letourneau_particle_2010}.
Particles and interactions are found in rule~146, and it is shown
that the intensity of traces in the average matrix is corresponding to~\cite{letourneau_particle_2010}. 
The differences of the classifications between this paper's and Wolfram's 
come from the different criterions. For example, in Wolfram class 2, some rules 
shows particle interactions and others not, which were classified into different 
classes in this paper.

Since the problems of storing and processing information can be
found in various fields, such as chemical systems~\cite{PhysRevLett.78.1190}
and hydrodynamics~\cite{perrard_wave-based_2016,harris_wavelike_2013},
and this method is not based on ECAs' specific features, so it is
potentially to be applied to other systems, such as birds flock~\cite{hildenbrandt_self-organized_2010},
traffic flow~\cite{nagatani_density_2000}, chaotic behaviors~\cite{perrard_wave-based_2016,harris_wavelike_2013},
and complex networks~\cite{brockmann_hidden_2013}. This method
can also be used to quantify the complexity of systems, for UTM was
considered having the highest complexity by~\cite{wolfram_new_2002},
which will make people have a deeper understanding of complex behaviors.

\section*{Acknowledgements}

The author is grateful for suggestions and assistance from Dr.~Lingfei~Wu
in University of Chicago, Dr.~Qianyuan~Tang in Nanjing University, Dr.~Kaiwen~Tian in University of Pennsylvania and Dr.~Hector~Zenil in Karolinska Institutet.



\appendix

\subsection{Particles in ECAs}

I define a local grid of cells in $M$ as a particle in ECAs. Backgrounds
are also particles, which do not have any interactions with other
particles or themselves. According to the definition of particles
in ECAs:

\begin{equation}
\mathfrak{P}=L_{t_{\text{max}}}(p_{L},p_{R})\,.
\end{equation}

In this study, the size of a space--time evolution is $(200,200)$. The
target particle

\begin{equation}
\label{particle-define-A}
\mathfrak{P}=L_{200}(100-2,100+2)\,.
\end{equation}

For the formula

\begin{equation}
p_{0}(p,n)=p(1-p)^{n-1}\,.
\end{equation}

\begin{figure}
\begin{centering}
\includegraphics[width=10cm]{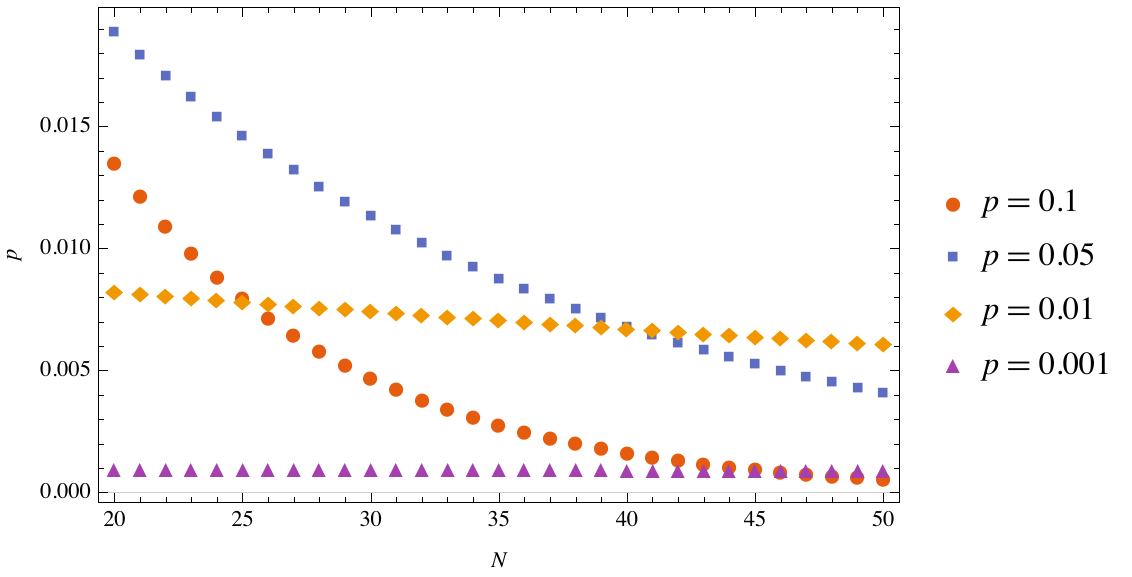}
\par\end{centering}

\protect\caption{\label{fig:Relation-of-}Relation of $p_{0}$ with different $p$
and $N$.}
\end{figure}

The number of $p$ is a priori hypothesis, choosing a proper $p$
will make images clear. Figure~\ref{fig:Relation-of-} shows that the
formula with different $p$ will not change its whole behavior. Experiments
show that choosing~$p=0.01$ will make average matrixes clear enough.

\subsection{Particles in Rule 146}

\begin{figure}
\begin{centering}
\includegraphics[width=8.5cm]{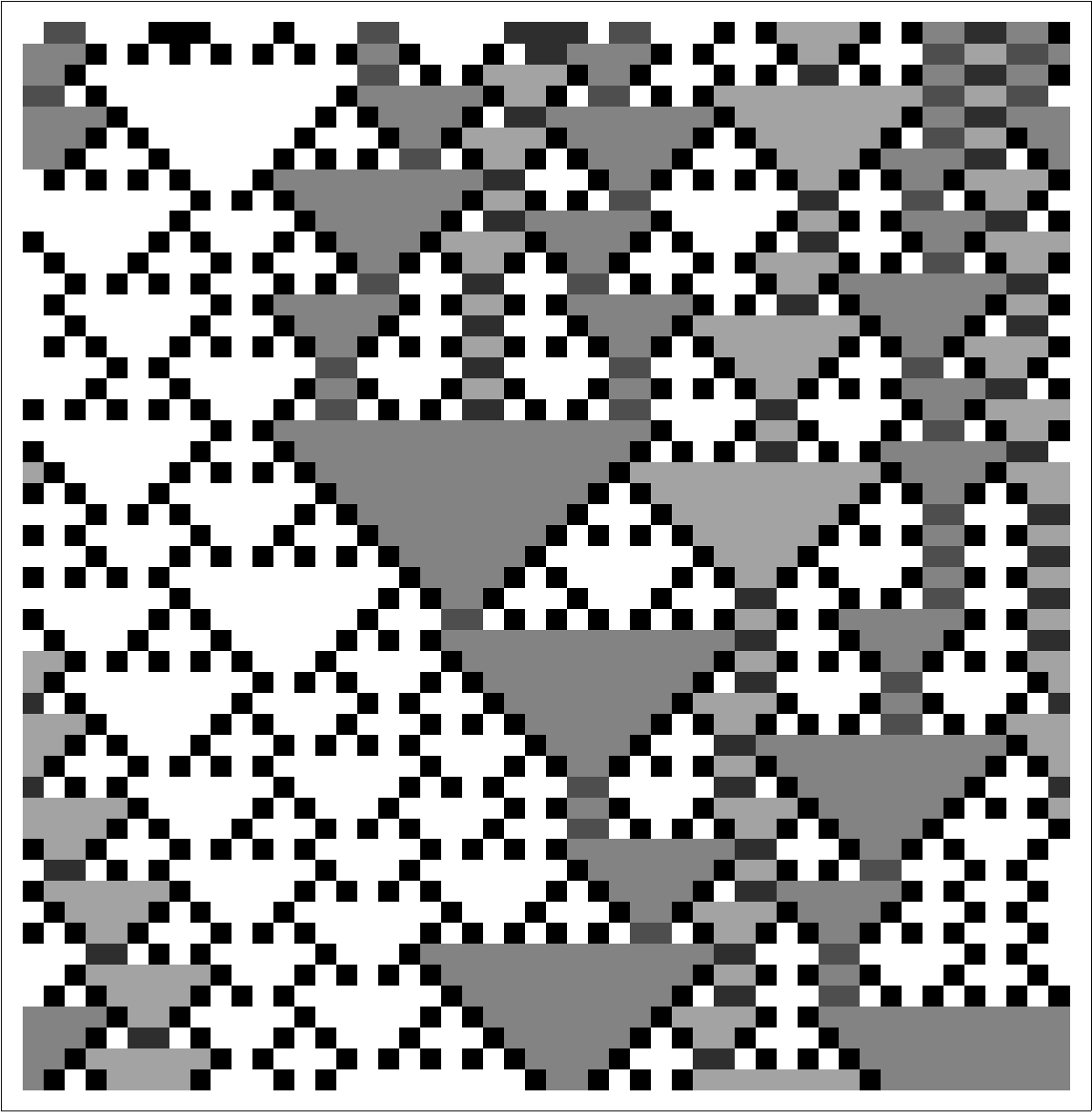}
\par\end{centering}

\protect\caption{\label{fig:Particles-146}Particles are found in rule~146.}
\end{figure}

Figure~\ref{fig:Particles-146} show particles in the space-time
for rule~146. These particles are also introduced by~\cite{letourneau_particle_2010}.

\subsection{Changes of lines' intensity}

The change of a line's intensity with time represents interactions between
particles. Because if a particle moves straight without any interactions,
the lines' intensity will remain unchanged through time. But if the
particle can be generated by other particles, it will not be found
before it was generated, so that the intensity of lines will change
through time, mostly, the intensity will get higher when $t$ is getting
higher. To get particles' changes of time, we define a function $D(t)$
to get paths' intensity:

\begin{equation}
D(t)=\max(L_{t})\,.
\end{equation}

\begin{figure}
\begin{centering}
\includegraphics[width=10cm]{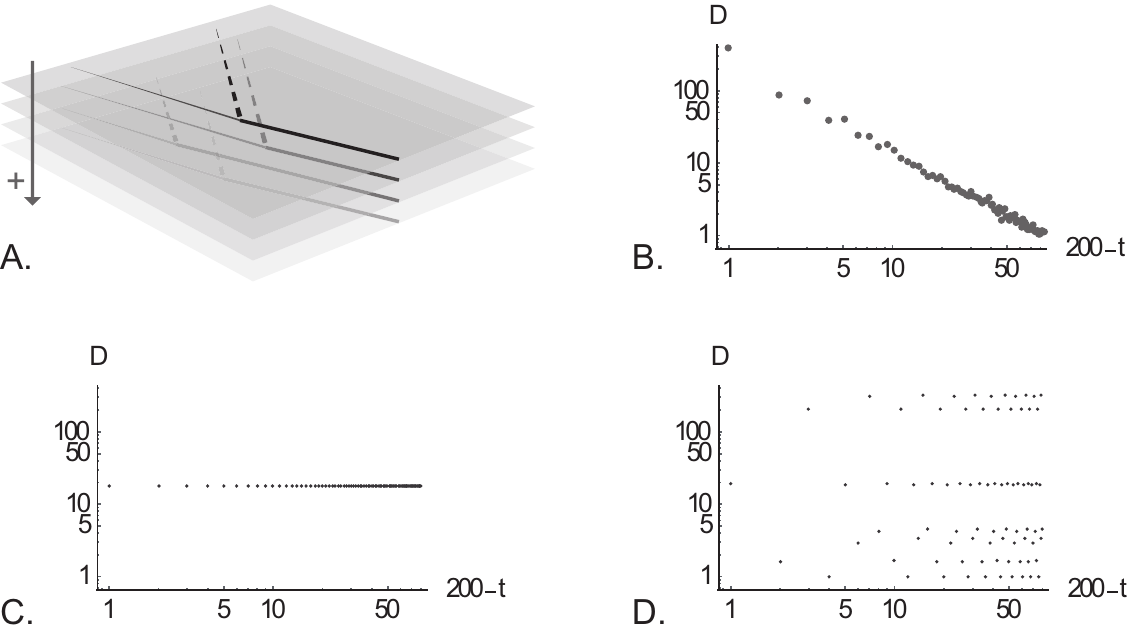}
\par\end{centering}

\protect\caption{\label{fig:Extracting-growth-of}Extracting growth pattern of particles.
A) The growth of particles' intensity represents interactions of particles.
B) An example of particle's intensity, generated with rule~149, which
has a growth pattern. C) Generated with rule~2, which do not has
growth pattern. D) Generated with rule~26, with multiple particles
and they all do not have growth pattern.}
\end{figure}

Figure~\ref{fig:Extracting-growth-of} shows the procedure of extracting
growth pattern of particles and three examples for rule~149, rule~2
and rule~26.

\subsubsection{The growth of particle traces' intensity for rule~146}

\begin{figure}
\begin{centering}
\includegraphics[width=8.5cm]{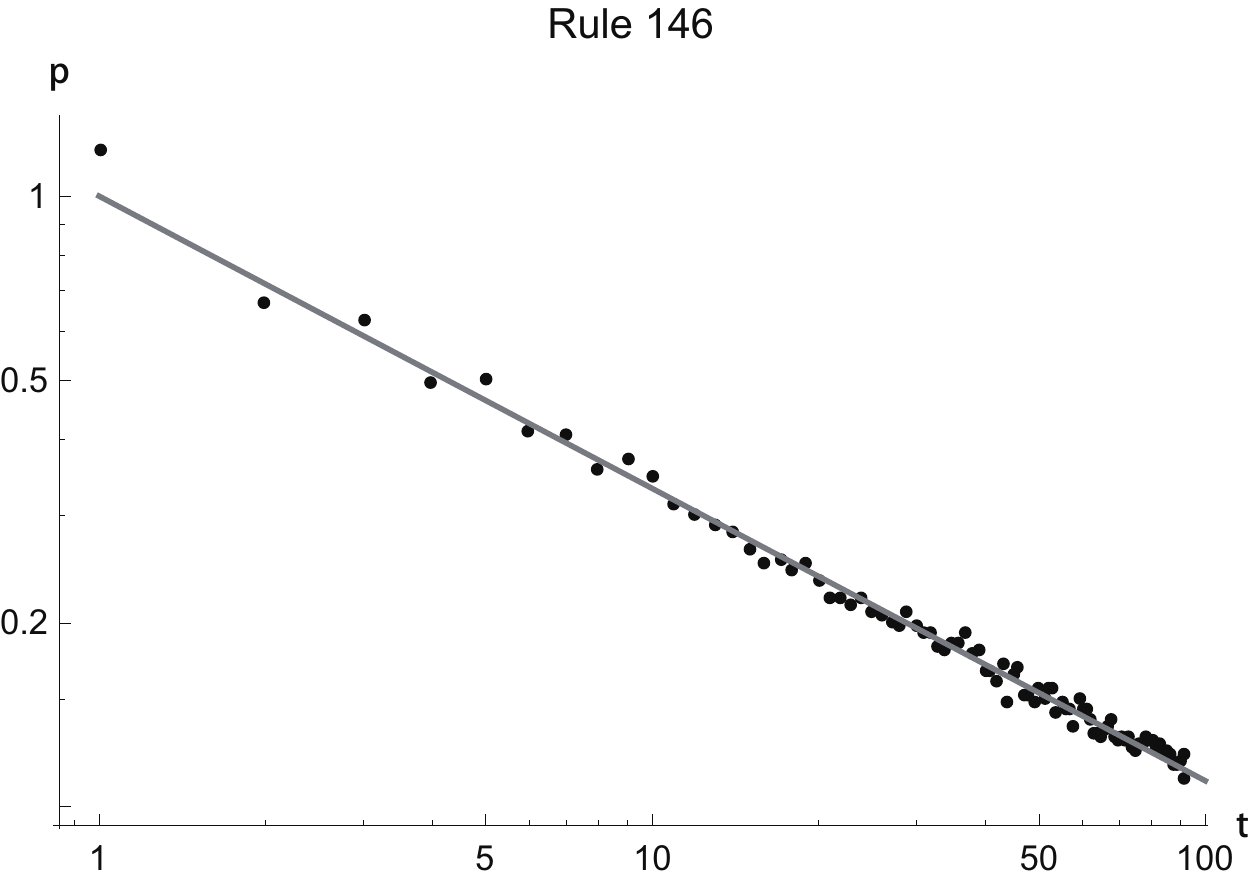}
\par\end{centering}

\protect\caption{\label{fig:146}The points are the data for the growth of particles'
traces. And the line is the figure of function $y=(t_{\textrm{max}}-t)^{-0.4789}$,
which has the same form as $n_{b}(t)\sim t^{-\alpha}$. It shows
that the power--law also is shown in this kind of measurement, and it
has a good fit when using the number of~$\alpha$ from~\cite{letourneau_particle_2010}.}
\end{figure}

Particles were found in Rule~146 (shown in Figure~\ref{fig:Particles-146}),
while also founded earlier in 2010~\cite{letourneau_particle_2010}.
In that study, the intensity of particles in rule~146 has a power-law
of the form

\begin{equation}
n_{b}(t)\sim t^{-\alpha}\,,
\end{equation}

\noindent with $\alpha=0.4789\pm0.0006$~\cite{letourneau_particle_2010}.
When this formula with this number was applied to the data in this
study (shown in Figure~\ref{fig:146}), it shows a good fit result.

\subsection{Typical Rules for Four Classes}

Some space--time evolutions of typical rules in each class shown in Table~\ref{tab:Typical-Rules-for}.

\begin{table}
\protect\caption{\label{tab:Typical-Rules-for}Typical Rules of Four Classes}

\centering{}\includegraphics[width=8.5cm]{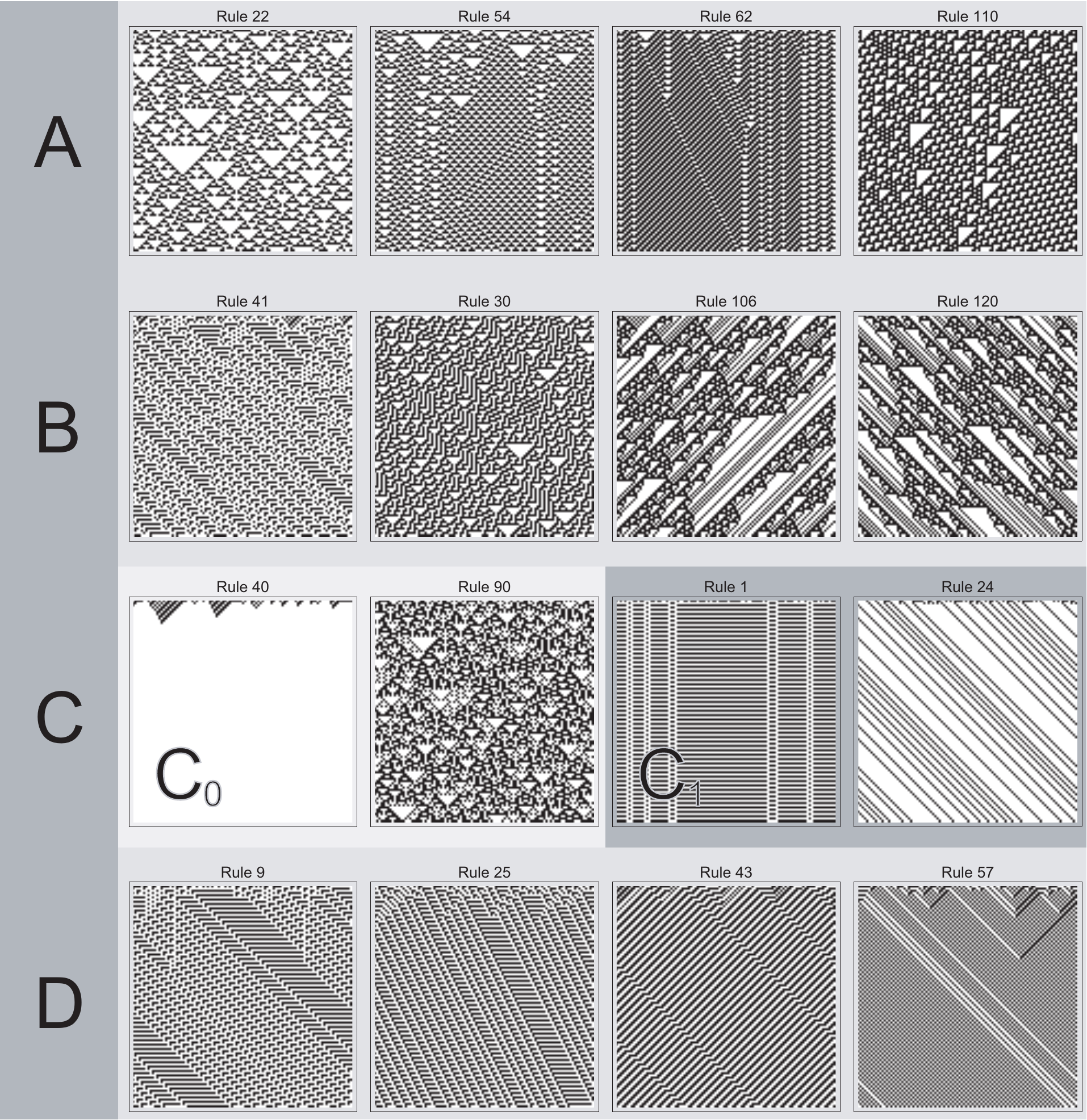}
\end{table}


\begin{thebibliography}{99}
\bibitem{banks_universality_1970}
Edwin~Roger~Banks, ``Universality in Cellular Automata,'' \textit{IEEE Annual Symposium (IEEE, 1970)}, pp.~194--215. doi:10.1109/SWAT.1970.27

\bibitem{cook_universality_2004}
Matthew~Cook, ``Universality in Elementary Cellular Automata,'' \textit{Complex Systems}, \textbf{15}(1), 2004 pp.~1--40.

\bibitem{wolfram_new_2002}
Stephen Wolfram, \textit{A New Kind of Science}, Champaign: Wolfram Media Inc, 2002.

\bibitem{note_CA_Class_2013}
Genaro J. Martinez, ``A Note on Elementary Cellular Automata Classification,'' \textit{Journal of Cellular Automata}, 2013 arXiv:1306.5577

\bibitem{asymptotic_in_CA_2013}
Hector~Zenil, and Villarreal-Zapata~Elena, ``Asymptotic Behavior and Ratios of Complexity in Cellular Automata,'' \textit{International Journal of Bifurcation and Chaos}, \textbf{23}(09), 2013 pp. 1350159.

\bibitem{compression_class_CA}
Hector~Zenil, ``Compression-based Investigation of the Dynamical Properties of Cellular Automata and Other Systems,'' \textit{arXiv preprint}, 2009 arXiv:0910.4042.

\bibitem{jakubowski_when_1996}
Mariusz~H.~Jakubowski, Ken~Steiglitz, and Richard~K.~Squier, ``When Can Solitons Compute?,'' \textit{Complex Systems}, \textbf{10}(1), 1996 pp.~1--22.

\bibitem{cross_boundary_Jurgen_Zenil_2015}
J{\"u}rgen~Riedel, Hector~Zenil, ``Cross-boundary Behavioural Reprogrammability Reveals Evidence of Pervasive Turing Universality,'' \textit{arXiv preprint}, 2017 arXiv:1510.01671

\bibitem{collision_based_computing_Adamatizky_2012}
Adamatzky, Andrew, and J{\'e}r{\^o}me Durand-Lose. ``Collision-based Computing,'' \textit{Handbook of Natural Computing}, Berlin Heidelberg: Springer, 2012 pp. 1949--1978.
``
\bibitem{cocke_universality_1964}
John~Cocke and Marvin~Minsky, ``Universality of Tag Systems with $P=2$,'' \textit{Journal of the ACM}, \textbf{11}(1), 1964 pp.~15--20, doi:10.1145/321203.321206

\bibitem{wolfram_statistical_1983}
Stephen Wolfram, ``Statistical Mechanics of Cellular Automata,'' \textit{Review of Modern Physics}, \textbf{55}(3), 1983 pp.~601--644, doi:10.1103/RevModPhys.55.601.

\bibitem{letourneau_particle_2010}
Paul-Jean~Letourneau, ``Particle Structures in Elementary Cellular Automaton Rule 146,'' \textit{Complex Systems}, \textbf{19}(2), 2010 pp.~143.

\bibitem{PhysRevLett.78.1190}
Marcelo~O.~Magnasco, ``Chemical Kinetics is Turing Universal,'' \textit{Physical Review Letters}, \textbf{78}(6), 1997 pp.~1190--1193, doi:10.1103/PhysRevLett.78.1190.

\bibitem{Asymptotic_Intrinsic_2016}
Hector~Zenil, J{\"u}rgen Riedel, ``Asymptotic Intrinsic Universality and Reprogrammability by Behavioural Emulation,'' \textit{arXiv preprint}, 2016 arXiv:1601.0033.

\bibitem{perrard_wave-based_2016}
St{\'e}phane~Perrard, Emmanuel~Fort, and Yves~Couder, ``Wave-Based Turing Machine: Time Reversal and Information Erasing,'' \textit{Physical Review Letters}, \textbf{117}(9), 2016, doi:10.1103/PhysRevLett.117.094502.

\bibitem{harris_wavelike_2013}
Daniel~M.~Harris, Julien~Moukhtar, Emmanuel~Fort, Yves~Couder, and John~W.~M.~Bush, ``Wavelike Statistics from Pilot-wave Dynamics in a Circular Corral,'' \textit{Physical Review E}, \textbf{88}(1), 2013, doi:10.1103/PhysRevE.88.011001.

\bibitem{hildenbrandt_self-organized_2010}
Hanno~Hildenbrandt, Cladio~Carere, and Charlotte~K.~Hemelrijk, ``Self-organized Aerial Displays of Thousands of Starlings: a Model,'' \textit{Behavioral Ecology}, \textbf{21}(6), 2010 pp.~1349--1359, doi:10.1093/beheco/arq149.

\bibitem{nagatani_density_2000}
Takashi~Nagatani, ``Density Waves in Traffic Flow,'' \textit{Physical Review E}, \textbf{61}(4), 2000, doi:10.1103/PhysRevE.61.3564.

\bibitem{brockmann_hidden_2013}
Dirk~Brockmann and Dirk~Helbing, ``The Hidden Geometry of Complex, Network-driven Contagion Phenomena,'' \textit{Science}, \textbf{342}(6164), 2013 pp.~1337--1342, doi:10.1126/science.1245200.

\end{thebibliography}
\end{document}